\documentclass[11pt]{article}

\usepackage{fontspec}
\usepackage{xeCJK}
\usepackage[margin=1in]{geometry}
\usepackage{graphicx}
\usepackage{booktabs}
\usepackage{array}
\usepackage{amsmath}
\usepackage{caption}
\usepackage[round]{natbib}
\usepackage[hidelinks]{hyperref}
\usepackage{xcolor}
\hypersetup{colorlinks=true, linkcolor=black, citecolor=blue!45!black, urlcolor=blue!45!black}

\title{\textbf{World Artificial Intelligence Cooperation Organization (WAICO):\\Mapping an Emerging Institution in the Global AI Governance Regime Complex}}

\author{
William Guey$^{1}$, Pierrick Bougault$^{1}$, Wei Zhang$^{1}$, Vitor D. de Moura$^{2}$, and Jos\'e O. Gomes$^{3}$ \\[0.5em]
\small $^{1}$Department of Industrial Engineering, Tsinghua University, Beijing, China \\
\small $^{2}$School of Social Sciences, Tsinghua University, Beijing, China \\
\small $^{3}$Department of Industrial Engineering, Federal University of Rio de Janeiro, Brazil
}
\date{Draft, June 2026}

\begin{document}
\maketitle

\begin{abstract}
\noindent Who sets the rules for artificial intelligence, and on what terms, has become a defining question of global governance. For several years that contest ran through principles and ethics codes; it now runs through institutions. China's proposed World Artificial Intelligence Cooperation Organization (WAICO; 世界人工智能合作组织) is the most consequential recent entrant and the least examined. We place WAICO within the emerging regime complex for AI and argue that its importance lies not in any single commitment but in the position it is designed to hold. Coding a cross-section of fifteen international AI governance instruments and institutions on how they admit members, how they are organized, and what they prioritize, we find that WAICO's proposed design joins three features that no constituted multilateral body currently combines: membership open to any sovereign state, no values or regime-type test for entry, and an agenda built around development and the global capability divide. The incumbent Western-led bodies gate membership by shared values and concentrate on rights and safety; the universal United Nations bodies are open but anchored in human rights; a development-first agenda is otherwise carried by the regional strategies of the Global South. Among constituted institutions, the only occupant of WAICO's intended position is China's own 2023 precursor initiative. We read this as the formation of a second, still-proposed pole in global AI governance, organized around sovereignty and development rather than rights and safety, and argue that WAICO would be the first standing organization built to anchor it. We report the full coding, state testable expectations against which the claim can be judged as the organization takes shape, and release the dataset for replication.

\vspace{0.6em}
\noindent\textbf{Keywords:} AI governance, regime complex, global governance of AI, WAICO, institutional design, Global South

\vspace{0.3em}
\noindent\textbf{关键词：} 世界人工智能合作组织，全球人工智能治理，人工智能治理，国际制度复合体，智能鸿沟，全球南方
\end{abstract}

\section{Introduction}
For most of its short history, the global governance of artificial intelligence has been written in the language of principles. Governments and international bodies issued ethics codes, declarations, and recommendations, and observers asked whether the world was converging on a shared set of values \citep{jobin2019global}. That phase is ending. Governance is now being built into institutions: standing bodies, treaties, summit processes, and expert panels, each with its own members, rules, and purpose \citep{schmitt2021mapping, geith2025emerging}. The decisive questions have shifted accordingly, from which principles states endorse to which institutions they join, and why.

It is into this institutional phase that China has proposed a new organization. Speaking at the opening of the World Artificial Intelligence Conference in Shanghai in July 2025, Premier Li Qiang announced, on behalf of the Chinese government, the creation of a World Artificial Intelligence Cooperation Organization (WAICO; 世界人工智能合作组织), to be headquartered in Shanghai, together with a thirteen-point Global AI Governance Action Plan (人工智能全球治理行动计划) \citep{govcn2025waic, mfa2025actionplan}. (WAICO, the proposed organization, should not be confused with the World Artificial Intelligence Conference, WAIC, the annual Shanghai meeting at which it was announced.) The organization is new, its design is still taking shape, and serious treatment of it has so far been confined to policy commentary \citep{basu2026pivot, nature2025china, cnn2025waic}.

This paper asks a more basic question than most commentary does. Rather than predict what WAICO will achieve, which no one can yet know, we ask where it sits. Every new institution enters a landscape that is already occupied, and its meaning comes as much from the gap it fills as from the words of its founders. Our claim is that WAICO is designed to occupy a position that no constituted global body currently holds, and that this position, more than any single line in its founding documents, explains why it is being created.

Because WAICO is still forming, it cannot be assessed through its conduct. It can, however, be placed. Every governance arrangement can be described by how it admits members, how it is organized, and what it prioritizes, and these choices fix its location in the field before it has a record to judge. We code a cross-section of fifteen AI governance instruments and institutions, WAICO among them, on exactly these dimensions, drawing each judgment from public documents, and use the result to locate WAICO relative to the bodies that already exist. Coding everything on the same terms, and reporting the full result in Table~\ref{tab:scores}, keeps the comparison open to inspection and lets any reader check, or contest, the placement we propose.

We make three contributions. First, we bring the mapping of AI governance institutions up to date, including the most recent entrants that earlier studies could not cover. Second, we add a reading of what these institutions stand for to what has mostly been a structural picture of how they relate. Third, we provide the first systematic placement of WAICO within the field. The picture that results is that the governance of AI is acquiring a second center of gravity, and that WAICO is the first organization proposed to anchor it.

\section{The global AI governance landscape: a short review}
Three strands of scholarship inform this paper. The first treats AI governance documents as data. In the most influential example, \citet{jobin2019global} examined eighty-four ethics guidelines and found that an apparent agreement on a few principles concealed wide disagreement over what those principles meant and how they should be applied. The lesson we carry forward is that surface consensus can hide real division, and that careful, comparable reading across documents is how the division becomes visible.

The second strand studies AI governance as a problem of institutions rather than texts. \citet{cihon2020fragmentation} asked whether international AI governance should be centralized in a single body or left fragmented across many, and concluded that fragmentation carries real costs but also real benefits, since competing forums can experiment and adapt. \citet{tallberg2023global} set out a research agenda for the field, distinguishing the empirical task of mapping and explaining global AI governance from the normative task of judging it. Building on that agenda, \citet{geith2025emerging} produced the first systematic analysis of what they call the AI regime complex, coding more than a hundred institutions and describing a field that is non-hierarchical, concentrated in Europe, and only weakly differentiated by function. Their work is the foundation we build on, and also the source of two gaps we try to fill: it maps how institutions relate rather than what they value, and its coverage closes before the newest and most consequential bodies appeared.

To interpret what follows we draw on three ideas from the study of international organizations. \citet{koremenos2001rational} argued that the design features of an institution, such as who may join and how decisions are made, are not arbitrary but respond to the kind of cooperation problem it is meant to solve. \citet{raustiala2004regime} introduced the regime complex, a set of overlapping and non-hierarchical institutions governing the same broad issue, now the standard way of describing AI governance. And \citet{morse2014contested} describe how dissatisfied states engage in \emph{competitive regime creation}, building new institutions to contest the rules of incumbents rather than working only within them. WAICO, we argue, is best read through this last lens.

The third strand concerns China specifically. \citet{roberts2021chinese} analyzed the policy, ethics, and regulation behind China's AI ambitions, showing that Beijing has long sought to be a rule-maker rather than a rule-taker in this domain. WAICO is a natural extension of that ambition, and recent commentary reads it as part of a broader effort to shape global AI norms in ways that foreground state sovereignty and appeal to developing countries \citep{basu2026pivot, nature2025china}. What has been missing is a placement of WAICO against the rest of the field. That is the gap this paper addresses.

\section{Data and methods}
\paragraph{Argument.} The field of AI governance is organized, above all, by two questions: who is allowed in, and what the institution is for. On the first, the established Western-led bodies are clubs: they admit members who share certain values, and they are built around managing the risks of advanced AI, namely safety, accountability, and the protection of rights. WAICO is presented as the mirror image. It is offered as open to any sovereign state, without a values test at the door, and it is organized not around risk and rules but around development: widening access to AI, building capacity in countries that lack it, and closing what Chinese statements call the global intelligence gap \citep{govcn2025waic, govcn2025waiczh, mfa2025actionplan, mfa2025actionplanzh, mfa2023initiative}. What makes this more than a contrast of slogans is that no constituted global institution already occupies WAICO's intended position, as the coding below shows.

\paragraph{Case selection.} We include a governance arrangement if it is (i) international or transnational in scope, (ii) the bearer of at least one explicit AI governance commitment, and (iii) among the most prominent instruments in its category, so that the four main blocs of the field, the Western and like-minded bodies, the universal multilateral bodies, the regional strategies of the Global South, and the China-led cluster, are each represented. This yields the fifteen cases in Table~\ref{tab:scores}. We do not aim at the full population; \citet{geith2025emerging} inventory more than a hundred bodies. The aim is a legible cross-section that spans the poles, and Section~6 returns to whether the central finding survives the bodies we leave out.

\paragraph{Coding.} Each case is coded from a public primary source (cited in Table~\ref{tab:scores}) on three groups of variables. \emph{Membership} is coded by the gate an institution places on entry: values-gated, capability-gated, regional, or universal-open. \emph{Formalization} follows the rational-design tradition of \citet{koremenos2001rational} and is scored from 0 to 5, one point each for a public charter, a secretariat, a budget, voting rules, and a defined membership. \emph{Normative orientation} is coded, in the presence-based manner of \citet{jobin2019global}, by the emphasis (0, 0.5, or 1) each instrument places on seven principle families; we summarize it in a single index,
\begin{equation}
\text{orientation} = (\text{sovereignty} + \text{development} + \text{openness}) - (\text{rights} + \text{safety}),
\end{equation}
which is negative on the rights-and-safety side and positive on the sovereignty-and-development side. Membership breadth is placed on a 0 (club) to 1 (universal) scale. One asymmetry deserves note: WAICO is read from a founding statement, while several comparators are read from negotiated legal texts. Early declarations are broad where mature law is specific. We address this by reading normative orientation from each body's constitutive statement of purpose, and by treating formalization (Figure~\ref{fig:def}) as a separate axis rather than letting it contaminate placement on Figure~\ref{fig:map}. WAICO is coded throughout from its stated, proposed design, and we do not assert its eventual membership; the reported regional spread of invitees is not yet an official roster, so we convert it into testable expectations (Section~7) rather than coding it.

\begin{table}[t]
\centering
\footnotesize
\caption{Coded values for the fifteen instruments and institutions. Orientation is the index in Eq.~(1); breadth runs 0 (club) to 1 (universal); development is the 0--2 emphasis on closing the global divide; formalization is the 0--5 score. WAICO is coded from its proposed design; all others are operating.}
\label{tab:scores}
\setlength{\tabcolsep}{5pt}
\resizebox{\textwidth}{!}{%
\begin{tabular}{lllrccc}
\toprule
Instrument / institution & Lead & Entry & Orient. & Breadth & Dev. & Form. \\
\midrule
WAICO (proposed) & China & universal-open & $+2.0$ & 1.00 & 2 & 1 \\
Global AI Governance Initiative 2023 & China & universal-open & $+1.5$ & 1.00 & 2 & 1 \\
BRICS Statement on AI governance & BRICS & club & $+2.0$ & 0.40 & 2 & 1 \\
African Union Continental AI Strategy & AU & regional & $+1.5$ & 0.33 & 2 & 2 \\
UNESCO Recommendation on AI Ethics & UNESCO & universal-open & $+1.0$ & 1.00 & 2 & 3 \\
UN Global Dialogue / Scientific Panel & UN & universal-open & $0.0$ & 1.00 & 2 & 3 \\
ASEAN Guide on AI Governance & ASEAN & regional & $0.0$ & 0.33 & 1 & 2 \\
AI Seoul Summit / Declaration & UK--Korea & club & $-0.5$ & 0.55 & 1 & 2 \\
AI Safety Summit / Bletchley & UK & club & $-1.0$ & 0.60 & 0 & 2 \\
Global Partnership on AI (GPAI) & West & values-gated & $-1.0$ & 0.50 & 1 & 4 \\
OECD AI Principles & OECD & capability-gated & $-1.0$ & 0.50 & 0 & 4 \\
Intl.\ Network of AI Safety Institutes & US & values-gated & $-1.0$ & 0.25 & 0 & 3 \\
G7 Hiroshima Process & G7 & values-gated & $-1.5$ & 0.30 & 0 & 3 \\
Council of Europe AI Convention & CoE & values-gated & $-1.5$ & 0.40 & 0 & 4 \\
EU AI Act & EU & regional & $-2.0$ & 0.33 & 0 & 5 \\
\bottomrule
\end{tabular}}

\smallskip
{\footnotesize Sources, by row: \citet{govcn2025waic}; \citet{mfa2023initiative}; \citet{brics2025ai}; \citet{au2024strategy}; \citet{unesco2021rec}; \citet{un2025dialogue}; \citet{asean2024guide}; \citet{seoul2024declaration}; \citet{uk2023bletchley}; \citet{gpai2020founding}; \citet{oecd2019principles}; \citet{aisi2024network}; \citet{g72023hiroshima}; \citet{coe2024convention}; \citet{eu2024aiact}.}
\end{table}

We proceed in two steps. We first read the institutions' founding language closely, to recover how they define who belongs and what they are for (Section~4). We then use the coding to test whether those self-descriptions translate into a structured pattern across the field (Section~5). The qualitative reading establishes the distinction; the coding shows it is general rather than anecdotal.

\section{Two logics of membership}
The clearest way to see what divides the field is to read how these bodies describe who belongs to them. The Global Partnership on Artificial Intelligence, launched in 2020 by a group of mostly advanced economies, defines itself around the responsible development of AI consistent with human rights and the shared democratic values of its members \citep{gpai2020founding}. Membership is, in effect, a declaration of political kinship: a state joins because it already holds a particular normative commitment. The same grammar runs through the bodies that cluster with GPAI. The OECD principles, the Council of Europe convention, and the G7 Hiroshima code each tie the governance of AI to human rights, democracy, and the rule of law \citep{oecd2019principles, coe2024convention, g72023hiroshima}. The club is defined by what its members share before AI enters the picture.

WAICO is described in a different grammar. In the Global AI Governance Action Plan and the earlier Global AI Governance Initiative, the organizing words are openness, equality, respect for sovereignty, mutual benefit, and development; the stated aim is to widen access and close the global intelligence gap, and belonging is framed around being a sovereign state rather than around shared values \citep{mfa2025actionplan, mfa2023initiative, govcn2025waic}. The Action Plan's own stated principles, in the official Chinese text, are benefiting the people, respect for sovereignty, a development orientation, safety and controllability, fairness and inclusiveness, and open cooperation \citep{mfa2025actionplanzh}; the center of gravity of that list is sovereignty, development, and openness rather than a test of shared values. Where the Western model treats common values as the price of admission, WAICO's proposed design treats their absence as a feature: a state need not subscribe to a political model to take part. This is the substance behind the claim that WAICO sets no values threshold at the door.

These are two theories of what an international AI institution is for. One builds a community of the like-minded and writes rules among its members, accepting a smaller membership as the price of cohesion. The other builds the widest coalition it can around access and development, accepting normative thinness as the price of breadth. \citet{roberts2021chinese} document China's long ambition to be a rule-maker rather than a rule-taker in AI; WAICO pursues that ambition not by beating the Western clubs on their own terms but by changing the terms, trading the values gate for an open door. The wager is that, for many states, an organization that asks only that they be sovereign will be more attractive than one that asks them to be the right kind of state.

If this reading is correct, it should show up not only in rhetoric but in the structure of the field: an institution's founding language and its measured position ought to line up. The next section shows that they do.

\section{Where WAICO sits}
\begin{figure}[t]
\centering
\includegraphics[width=0.92\textwidth]{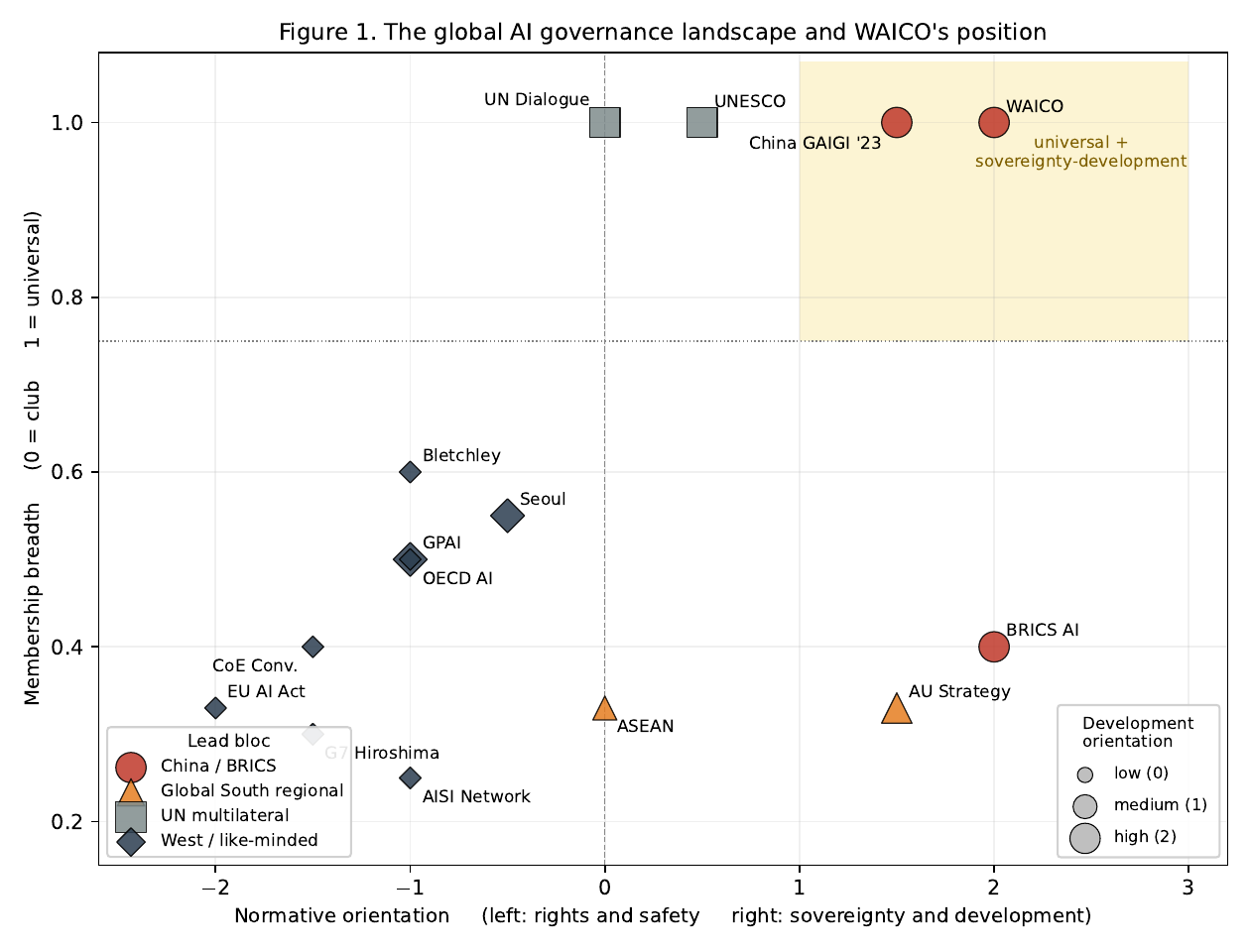}
\caption{The global AI governance landscape and WAICO's position. Horizontal axis: normative orientation (left, rights and safety; right, sovereignty and development). Vertical axis: membership breadth (0, club; 1, universal). Marker size reflects development emphasis. The shaded area, combining universal membership with a sovereignty-and-development orientation, is occupied only by WAICO and China's own 2023 initiative.}
\label{fig:map}
\end{figure}

Figure~\ref{fig:map} places the fifteen cases on two axes. The horizontal axis is normative orientation, the index of Eq.~(1), running from rights and safety on the left to sovereignty and development on the right. The vertical axis is membership breadth, from narrow clubs at the bottom to universal openness at the top. Marker size reflects development emphasis. The resulting pattern is clear.

The Western and like-minded bodies cluster on the left, at low to middling breadth: the Global Partnership on AI, the OECD principles, the EU AI Act, the Council of Europe convention, the G7 Hiroshima process, the Bletchley and Seoul declarations, and the network of AI safety institutes, grouped around rights and safety and around membership that is values-based or regional. The universal bodies of the United Nations and UNESCO sit at the top, open to all, but near the center of the horizontal axis rather than at its development end, because they pair a development agenda with an anchoring in human rights. The Global South's own strategies and the BRICS statement sit on the right, sharing WAICO's development orientation, but low down, because they are regional or club-based.

The upper-right region, where universal openness meets a sovereignty-and-development orientation, is nearly empty. Only two of the fifteen cases fall there: WAICO and China's own 2023 Global AI Governance Initiative, a declaration rather than an organization, which WAICO is meant to institutionalize. Among constituted multilateral bodies, then, the cell is empty; WAICO's only neighbor is the statement of intent it is built to turn into an organization, and its nearest measured neighbors beyond that are other Global South and China-aligned instruments. The map also shows why the membership rule carries so much weight: across the set, the bodies that screen members by shared values are the same ones that lean toward rights and safety, while the bodies open to all without a values test are the ones more willing to foreground sovereignty and development. WAICO's proposed design joins the open door to the development agenda, and that pairing is what sets it apart.

\section{Does the niche survive the bodies we left out?}
A natural objection is that the niche looks empty only because development-oriented universal bodies were excluded. The clearest candidate is the International Telecommunication Union, whose mandate centers on closing the digital divide and whose membership is universal; the G20, the UN Global Digital Compact, and the multilateral development banks raise the same question. Two considerations bound this concern. First, these are functional or technical agencies operating within the United Nations framework, not general AI rule-making bodies, and they do not adopt a value-neutral, sovereignty-first membership logic; on our axes they would sit with the universal bodies near the center, not in the sovereignty-and-development corner. Second, and more important, the claim we defend is not that universal development work is unprecedented, which it plainly is not, but that no existing body combines universal, value-neutral membership with a development-first agenda as a general-purpose AI governance organization. That combination, not development activity as such, is what is currently unoccupied. The point is narrow but, we think, real.

\section{What kind of institution it is}
\begin{figure}[t]
\centering
\includegraphics[width=0.86\textwidth]{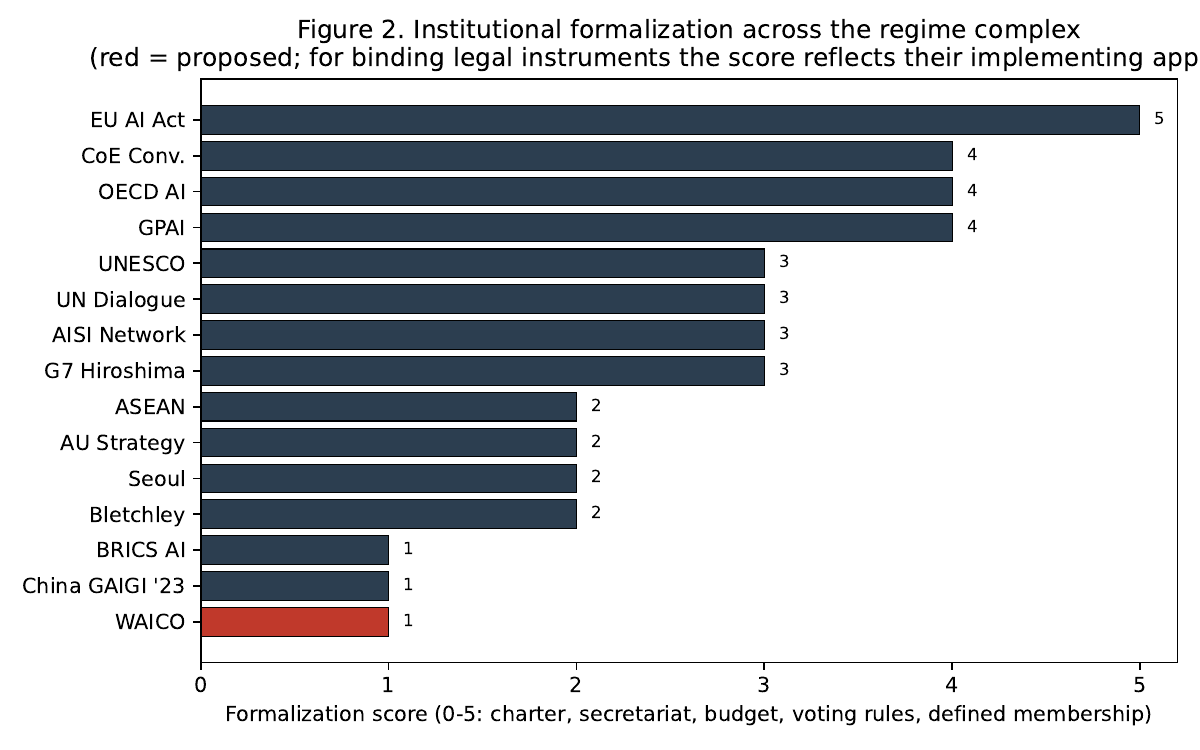}
\caption{Institutional formalization, scored on whether each case has a charter, a secretariat, a budget, decision rules, and a defined membership. WAICO, shown in a different shade as the only proposed body, sits at the bottom of the set. For binding legal instruments the score reflects their implementing apparatus.}
\label{fig:def}
\end{figure}

A distinctive position is not the same as a finished institution, and here WAICO stands out for the opposite reason. Figure~\ref{fig:def} scores each case on how much of its machinery is in place: a charter, a secretariat, a budget, decision rules, and a defined membership. WAICO sits at the bottom of the set. At present it has a name, a host city, and a statement of purpose, but the apparatus that would let it act has not yet been specified. The most formalized cases are, by contrast, the binding or club-based Western instruments, with the EU AI Act the most complete; for a binding instrument the score reflects an established implementing apparatus rather than an organizational membership, which is why the axis measures formalization rather than organizational maturity alone.

This deserves to be stated as a finding, not a footnote. WAICO enters the field with the strongest claim to a distinctive position and the least developed means of acting on it. The distance between an ambitious purpose and an unspecified organization is exactly where the hard questions live, and it is where skeptical observers have pointed: who pays, who decides, and how the promise of openness is to be reconciled with effective rule-making all remain undefined \citep{basu2026pivot, nature2025china}. The thinness of WAICO's current design is therefore not a reason to dismiss it, but the main thing to watch as it develops.

\section{The development divide WAICO names}
WAICO's stated reason for existing is to help countries that have been left behind by the AI revolution. It is fair to ask whether the gap it names is real in the institutions that already exist. The coding suggests it is. Among the fifteen cases, none of the eight Western and like-minded bodies places development at the center of its agenda: on the development scale none scores high, and most score zero. Of the seven others, six treat closing the divide as a central goal, scoring high, and the seventh, the ASEAN guide, places a medium emphasis on it. This does not mean the universal bodies treat development as an afterthought: the United Nations and UNESCO carry a substantial development agenda alongside their human-rights anchoring. The point is one of emphasis. Where the China-led, BRICS, African Union, and, to a lesser degree, ASEAN instruments foreground development, the most formalized Western clubs treat it, at most, as a secondary concern.

This matters for how we read WAICO. It does not show that WAICO will close the divide, which depends on who joins and on what resources follow. But it shows that the space WAICO is moving into is genuinely open. The most powerful and best-organized institutions in AI governance have, for the most part, not made the global distribution of AI capacity their central business. WAICO proposes to make it the purpose of a global organization. Whether or not one welcomes that proposal, the opening it responds to is visible in the existing institutional record.

\section{Reading WAICO: complement, competitor, or coordinator}
What, then, is WAICO likely to be? Three readings are consistent with its design, and the evidence favors some more than others.

As a \emph{competitor} to the Western clubs, WAICO is well placed. It stands on the opposite side of the field, and it removes the values test that keeps much of the world outside bodies like the Global Partnership on AI. This is competitive regime creation in the sense of \citet{morse2014contested}: a dissatisfied coalition builds a new institution to contest the terms of the incumbents. For states that have felt like guests rather than members in the existing system, an organization that asks only that they be sovereign is an attractive alternative.

As a \emph{complement} to the United Nations, WAICO is plausible but not assured. The two share an open door, and Chinese statements are careful to present the United Nations as the main venue for global rules \citep{basu2026pivot}. Yet WAICO's emphasis on sovereignty and development is not the same as the United Nations' emphasis on rights, and the two could compete for the loyalty of developing countries as easily as they could reinforce one another.

As a \emph{coordination layer} for the Global South, WAICO fits the evidence best. Its closest neighbors in the field are Chinese and Global South instruments, and its orientation is toward capacity-building and access rather than rule-enforcement. An organization that gathers developing countries, builds a common position, and then carries that position into broader forums is a coherent reading of what the design points toward. On this view WAICO is less a rival rule-book than a caucus with an address.

Across all three readings, one conclusion holds. The field of AI governance is no longer organized around a single center. A second pole is forming, defined by open membership and a development-first agenda, and WAICO is the first organization proposed to anchor it. Whether that pole hardens into a lasting division or is absorbed into United Nations-centered cooperation is the central open question, and one that the coming years, not this paper, will settle.

\section{What to watch}
Because the argument is about the position WAICO is built to hold, it can be tested as the organization takes concrete form. We state three expectations that follow from the reading offered here, to be judged against the public record as it appears.

\begin{enumerate}
\item[\textbf{1.}] WAICO's membership will lean toward developing countries that are largely absent from the Western-led bodies, rather than toward the advanced economies that dominate those bodies.
\item[\textbf{2.}] WAICO's founding arrangements will not impose a values-based or political condition on membership, consistent with the open-door positioning described here.
\item[\textbf{3.}] There will be little overlap between WAICO's membership and that of the Global Partnership on AI. Because WAICO is proposed as open to all, this would reflect self-selection by states on opposite sides of the field rather than any formal exclusion.
\end{enumerate}

These expectations follow directly from where WAICO sits in Table~\ref{tab:scores} and Figure~\ref{fig:map}. If they hold, they corroborate the reading that WAICO is built to anchor a second, development-oriented pole. If they do not, the reading should be revised. The released dataset is arranged so that WAICO's details can be added as they become public and these expectations evaluated without redoing the rest of the work.

\section{Limitations}
Four limitations should be kept in view. First, the coding applies structured judgment to public documents, and reasonable readers may score individual cases differently; the full scores appear in Table~\ref{tab:scores} and the dataset is released so that codings can be inspected, contested, and refined, but formal inter-coder reliability testing is a natural next step. Second, the two axes are theory-driven, following the membership dimension of \citet{koremenos2001rational} and the principle-coding of \citet{jobin2019global}, rather than extracted from the data; they are an imposed analytical frame, chosen because they correspond to the two questions, who may join and what the body is for, that organize the field. Third, WAICO is read from its stated intentions, and stated intentions can differ from what an organization eventually becomes; the expectations above are the means of testing that gap rather than assuming it away. Fourth, the fifteen cases are a cross-section spanning the field's main poles, not the full population; Section~6 argues that the central finding survives the most likely omitted occupants, but extending the coding to the full inventory of \citet{geith2025emerging} would test this at scale. None of these qualifications touches the core, design-level claim, which rests on public sources and on the position WAICO has chosen for itself.

\section{Conclusion}
WAICO is most usefully read not as an outcome but as a move into an already crowded field. Placed against fifteen existing instruments and institutions, it is designed for a position that no constituted global body currently holds: open to all, free of a values test, and centered on development rather than on rules and risk. That position, rather than any one phrase in its founding documents, is the clearest explanation of why it is being created and the best guide to what it might become. The governance of AI is gaining a second center of gravity, and WAICO is the first organization proposed to sit at its core. As the organization takes shape, the dataset and the expectations offered here are meant to be revisited, so that a reading made at the moment of its proposal can be tested against what it turns out to be.

\section*{Data and code availability}
The coded dataset, the analysis script, and the figure source are openly available at \url{https://github.com/williamguey/waico-ai-governance}. Each case is coded from a public primary source, whose address is recorded with the data and cited in Table~\ref{tab:scores}. The dataset is structured to accept WAICO's details, as they become public, for evaluation of the expectations set out above.

\section*{Competing interests}
The authors declare no competing interests.

\bibliography{references}

\end{document}